\documentclass{osa-article}

\journal{oe}

\begin{document}

\title{Frequency-doubled Nd:YAG MOPA laser system with programmable rectangular pulses up to 200 microseconds}

\author{Maximilian Beyer,\authormark{1,2,6,*} Jack C. Roth,\authormark{1,3,6} Eustace Edwards,\authormark{1,4} and David DeMille \authormark{5,$\dagger$}}

\address{\authormark{1}Department of Physics, Yale University, New Haven, CT 06520, USA \\
\authormark{2}Currently: Department of Physics and Astronomy, Vrije Universiteit Amsterdam, 1081 HV Amsterdam, The Netherlands\\
\authormark{3}Currently: Department of Physics, University of California, Berkeley, Berkeley, CA 94720, USA\\
\authormark{4}Currently: Raytheon Technologies, 235 Presidential Way, Woburn, MA 01801 \\
\authormark{5}Department of Physics, University of Chicago, Chicago, IL 60637, USA\\
\authormark{6} These authors contributed equally to this work.}

\email{\authormark{*}mbeyer@mbeyer.ch } %% email address is required
\email{\authormark{$\dagger$}ddemille@uchicago.edu} %% email address is required

%%%%%%%%%%%%%%%%%%% abstract %%%%%%%%%%%%%%%%

\begin{abstract} 
A compact frequency-doubled diode-pumped Nd:YAG master-oscillator power-amplifier laser system with programmable microsecond pulse length has been developed. Analog pulse shaping of the output from a single-frequency continuous-wave Nd:YAG oscillator, and subsequent amplification, allowed the generation of rectangular pulses with pulse lengths on the order of the Nd:YAG fluorescence lifetime. 
Temporally flat-top pulses of 1064~nm light with 520~mJ pulse energy, 2.6~kW peak power, and 200~$\mu$s duration, with linewidth below 10~kHz, were obtained at a repetition rate of 2~Hz. Second harmonic generation in a LBO crystal yielded pulses of 262~mJ and 1.3~kW peak power at 532~nm. The peak power can be maintained within 2.9\% over the duration of the laser pulse, and long-term intensity stability of 1.1\% was observed. The spatially flat-top beam at 1064 nm used in the amplifier is converted to a Gaussian beam at 532~nm with beam quality factor $M^2=1.41(14)$ during the second harmonic generation. This system has potential as a pump source for Ti:sapphire, dye, or optical parametric amplifiers to generate tunable high-power single-frequency radiation for applications in precision measurements and laser slowing. 
\end{abstract}

%%%%%%%%%%%%%%%%%%%%%%%%%%  body  %%%%%%%%%%%%%%%%%%%%%%%%%%
\section{Introduction}

High-power broadly tunable laser sources with ${\sim}100$ microsecond pulse duration and narrow linewidth are desired for a wide range of applications. In atomic and molecular physics, potential applications include optical Stark deceleration \cite{fulton06a,coppendale11a} with near resonant light and precision measurements requiring wavelengths only accessible via nonlinear processes \cite{seiler05a, haase15a}. These sources could also find application as Bragg diffraction lasers in matter-wave interferometry \cite{muller08a}. However, to the best of our knowledge, lasers of this type have not yet been developed. A plausible way to obtain these characteristics would be to use a high power, long-pulse pump laser to amplify a lower power, broadly tunable narrow-linewidth laser. Existing high power systems are restricted to laser media like Nd:YAG with limited tunability and pulse durations significantly shorter than the fluorescence lifetime of about 230~$\mu$s \cite{koechner06a}.

To illustrate the possible utility of such a tunable, high power, long-pulse laser system, we focus on a particular application: laser slowing and cooling of molecules. Laser slowing and cooling is typically accomplished via forces from near-resonant photon scattering, which is associated with spontaneous emission \cite{metcalf99a}. However, with sufficiently intense laser beams, stimulated light forces can be used to dramatically increase slowing and cooling rates, despite a lower scattering rate \cite{soding97a}.  Such schemes are particularly attractive \cite{chieda11a, aldridge16a, yin18a, yang16a, kozyryev18a} for laser slowing and cooling of molecules, which can have optically accessible electronic transitions but suffer from optical pumping to dark states each time a photon is scattered \cite{di-rosa04a, kozyryev17a}. Although laser slowing, cooling, and trapping of molecules has been demonstrated using photon scattering forces \cite{shuman09a, shuman10a, barry14a, zhelyazkova14a, hemmerling16a, hummon13a, baum2020a}, the required long slowing distance required and very modest cooling rates deliver orders of magnitude smaller fluxes of slow molecules than are typical in atomic experiments.  

Promising experimental implementations of stimulated-force slowing and/or cooling, using the particular approach of the so-called bichromatic force (BCF) \cite{grimm94a,soding97a}, have been reported for atoms \cite{soding97a, partlow04a, chieda12a} and for molecules \cite{galica18a, kozyryev18a, wenz20a}. In each of these experiments, the use of continuous wave laser sources limited the attainable power; hence, the intensity required for strong slowing and cooling forces could be applied only over a small fraction of the area occupied by the atomic or molecular beam. High-energy pulsed laser sources offer the possibility to apply substantial forces over a large area, though over shorter times. This would be sufficient for experiments using pulses of atoms and molecules, for example molecules produced in the ablation-loaded cryogenic buffer gas beam sources \cite{hutzler12a} used in all molecular laser cooling experiments to date. Here, a laser with intensity of $\sim$1 kW/cm$^2$, tuned near resonance, could in principle slow molecules with a typical cryogenic beam velocity of $\sim$200~m/s, over an area of $\sim$1~cm$^2$, to a near standstill within a single pulse of $\sim 100~\mu$s (corresponding to a molecular travel distance of $\sim$1 cm). This efficient process would enable trapping and further cooling of much larger molecular samples; it could also be applied to a very wide range of molecular species, due to the reduced number of spontaneous emission events required during slowing. The stimulated force depends strongly on the ratio of Rabi frequency to detuning \cite{chieda12a} and for this reason temporally flat-top laser pulses are needed for optimal BCF performance.

With this motivation, we report here on a compact, frequency-doubled diode-pumped Nd:YAG master oscillator power amplifier (MOPA) laser system with programmable pulse length up to $\sim$200 microseconds, good beam quality, programmable temporal intensity profiles, and narrow linewidth. This may enable the amplification of low power tunable lasers in a gain medium like Ti:sapphire, or via optical parametric processes. Compared to previously reported Nd:YAG MOPA systems \cite{meijer17a,yarrow07a,xu15a,bian16a,bian16b, guo16a, zong18a, zong19b}, we obtain longer temporally flat-top pulses and a narrower linewidth while reducing the complexity of the optical setup, which will facilitate the integration in existing experiments.

\section{Experimental setup}

\begin{figure}[h!]
\centering\includegraphics[width=0.8\columnwidth]{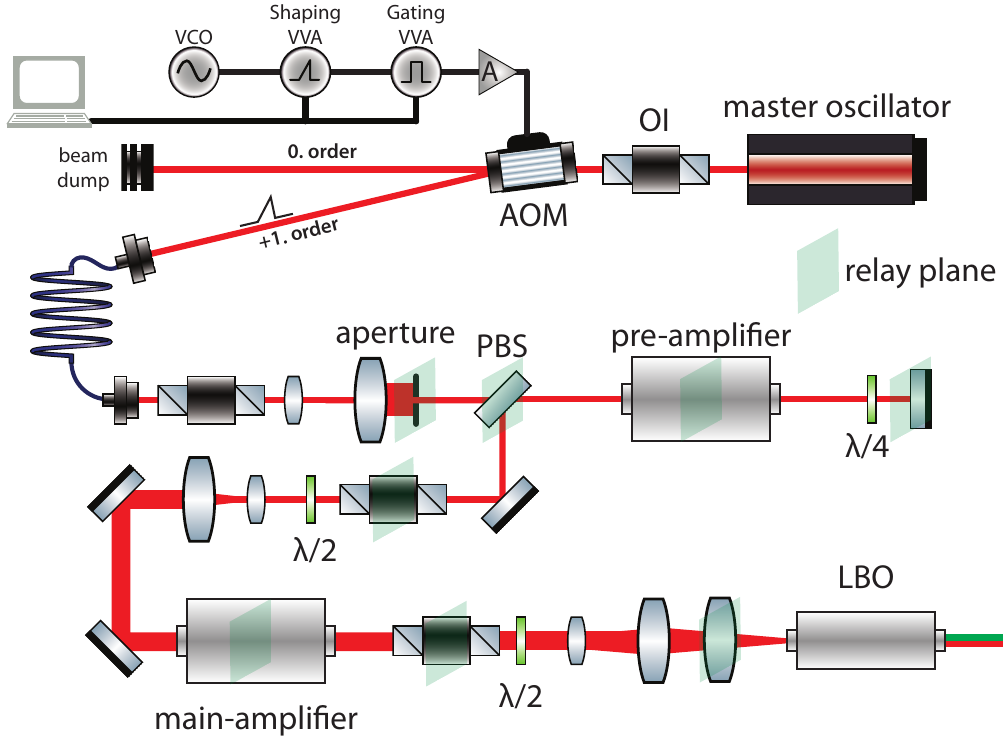}
\caption{Schematic diagram of the frequency-doubled Nd:YAG MOPA laser system. The spatial flat-top beam profile is maintained throughout the laser system using relay lenses (not shown) with the image planes indicated in green. VCO: Voltage-controlled oscillator. VVA: Voltage-variable attenuator. AOM: Acousto-optic modulator. OI: Optical isolator, PBS: Polarizing beam splitter. $\lambda/4$: Quarter-wave plate. $\lambda/2$: Half-wave plate. LBO: nonlinear crystal for second harmonic generation. }
\label{fig_layout}
\end{figure}

The optical layout of the laser system is illustrated in Fig.~\ref{fig_layout}. The entire system is compact (about 1m$^2$). The master oscillator is a modified continuous-wave (cw) nonplanar ring oscillator laser \cite{lumentum}, providing up to 400~mW output power at the fundamental wavelength of 1064~nm with a specified linewidth below 5~kHz. The beam from this laser is passed through an amplitude-modulated acousto-optic modulator (AOM), driven at frequency 80~MHz, to produce a seed laser pulse. A voltage-controlled oscillator (VCO) provides the 80~MHz RF signal, which is passed through two voltage-variable attenuators (VVA) \cite{vva}. The first VVA with a rise time of 8~ns and an extinction ratio of 28~dB is used for pulse shaping of the 80~MHz carrier envelope. To achieve a higher on/off extinction ratio, the shaped RF pulse is gated using a second, slower VVA (rise time 14~$\mu$s, extinction ratio 40~dB). The control voltage for the attenuators is provided by a NI-DAQ card (PCIe-6353) connected to a computer, which allows for interactive pulse shaping by monitoring the time profile of the laser pulses at various stages of the laser system using photodiodes. Enough flexibility in the voltage time profile was obtained by defining it through ten evenly spaced points and using linear interpolation in between.  The first-order diffracted beam from the AOM is coupled into a polarization maintaining single mode fiber and delivered to the amplifier system, which consists of one pre-amplifier and one main-amplifier Nd:YAG gain module \cite{northrop}. A double-pass configuration is used for the pre-amplifier. This employs a retroreflection with polarization rotated 90$^\circ$ to avoid standing waves in the Nd:YAG module, and a polarizing plate beamsplitter to separate the incoming and outgoing beams \cite{lee96a}. A quarter-waveplate was used to rotate the polarization in the double-pass configuration, after making sure that it showed the same performance as a Faraday rotator \cite{fluck2000a}. We found no sign of thermally induced birefringence on the beam profile or with regard to parastic self-lasing of the amplifier.

Both gain modules are diode pumped at 808~nm and are operated at a maximum current of 175~A. Diode drive current pulse lengths are 750~$\mu$s and 500~$\mu$s, for the pre-amplifier and main-amplifier, respectively (see Fig.~\ref{fig_time_profile}). The system operates at 2~Hz repetition rate, which is motivated by its future application for laser cooling a cryogenic molecular beam. A lower repetition rate also allows for longer diode pump lengths to enable a higher gain and the generation of temporally flat-top  pulses with durations similar or longer than the fluorescence lifetime of the Nd:YAG crystal, while staying below the specified duty cycle of 0.5\%.  

To increase the extraction efficiency from the amplifier modules, the seed laser spatial mode is converted from a Gaussian to nominally flat-top beam; this enables higher filling of the gain medium, while avoiding diffraction. The flat-top profile is generated by expanding the seed beam to 2~mm $1/e^2$ diameter and passing it through an aperture 1~mm in diameter. It is then propagated through the amplifier system using a series of relay lenses with the relay planes indicated in Fig.~\ref{fig_layout}. The relay lenses are also used to change the beam diameter to ensure a filling (with respect to the diameter of the gain medium) of 87\% and 69\% in the pre- and main-amplifiers, respectively. 
The obtained output power did not decrease when reducing the seed power by 30\%. 

To reduce parasitic self-lasing from back-reflections, optical isolators are inserted after the fiber outcoupler that directs the seed beam to the pre-amplifier, and between the pre-amplifier and the main-amplifier. Due to the compact size of the amplifier system (1~m x 1~m) and the long pump diode pulses, a fraction of the spontaneous emission fluorescence can easily be amplified leading to self-lasing in the system which spoils the temporally flat-top profile and can lead to optical damage. To prevent this effect, additional apertures are placed at the focal points of the relay lenses to reduce the amount of fluorescence propagating through the system.

To generate the second harmonic (SH) at 532~nm, the 1064~nm beam is focused into a temperature-stabilized $l=50$~mm long LBO crystal with AR coating for 1064~nm and 532~nm, held at 149.6~C for non-critical phase matching.  We use a beam waist of $\omega=70~\mu$m at the center of the crystal medium and a focusing parameter \cite{boyd68a} of $\xi=l\lambda/2\pi n\omega^2=0.95$. 

\section{Results and discussion}

\begin{figure}[h!]
\centering\includegraphics[width=\columnwidth]{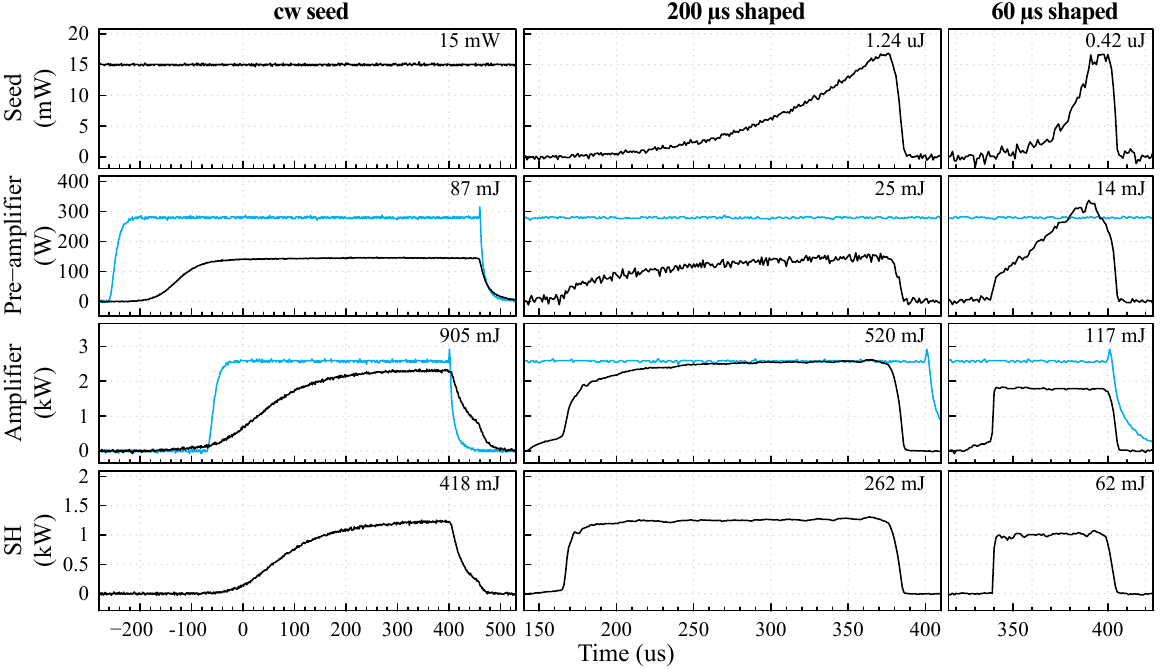}
\caption{Temporal pulse shapes and instantaneous power of the laser system for an unshaped cw seed laser (left column), and for seed laser pulses shaped to obtain rectangular second harmonic pulses with length 200~$\mu$s (middle column) and 60~$\mu$s (right column). From the top to bottom row: time profiles of the seed, the double-pass pre-amplifier output, the main-amplifier output, and the second harmonic (SH) pulses. The pulse energies are given in each panel. The blue lines in the second and third rows indicate the time profile of the current driving the pre-amplifier and main-amplifier pump diodes, respectively. The same drive currents are used for each pulse.  Note the change in time scales between columns. }
\label{fig_time_profile}
\end{figure}

Fig.~\ref{fig_time_profile} demonstrates the pulse shaping capabilities of the MOPA laser system. The first row depicts the time profile of the amplifier seed, generated by amplitude modulation of the master oscillator output using the AOM. The second and third rows depict the pulse time profile after a double-pass through the pre-amplifier module and single pass through the main-amplifier module, respectively. 
The last row displays the time profile of the SH. To avoid self-lasing caused by amplifying fluorescence from the pre-amplifier in the second gain module, we reduce the pump length in the main-amplifier. The upper limit for the pump length of the amplified 1064~nm pulse is related to the pulse length of the pump diodes. The amplified pulse is shorter than the pump diode current pulse because of the build up time of 100~$\mu$s, illustrated in the cw seed column of panels. As can be seen in the left column of Fig.~\ref{fig_time_profile}, after roughly 100~$\mu$s for the pre-amplifier and 300~$\mu$s for the amplifier, a steady state is reached that leads to a flat time profile. This indicates an equilibrium between populating the upper level of the Nd:YAG gain medium using pump diodes and depleting it by stimulated emission. A small pedestal at the beginning of the amplified fundamental pulse is caused by the smaller extinction ratio of the shaping VVA, while the gating VVA is switched. The nonlinear second harmonic generation process strongly suppresses the pedestal and further shortens the shape of the 532~nm pulse compared to the 1064~nm pulse.

We optimized our system to obtain a flat time profile in the SH pulse. To accomplish this, the seed time profile must anticipate the time dependent gain of the amplifier system and the non-linearity of the second harmonic generation \cite{yarrow07a, meijer17a}. The gain modules are driven using the same currents shown in the cw panel, with a pump time much longer than the target pulse length. This is done in order to maximize the population inversion in the gain medium, and therefore to maximize the gain, whenever the seed pulse is on. The population inversion $\Delta$ is maximal at the beginning of the seed pulse, then is rapidly depleted as the pulse stimulates emission while propagating through the gain medium. For pulse energies much smaller than the stored energy of the amplifier, the gain $G$ is an exponential function of the population inversion $G=\exp(\Delta\sigma L)$ where $\sigma$ is the absorption cross section ($28\times10^{-20}$~cm$^2$ for Nd:YAG) and $L$ is the length of the amplification medium (63~mm for the pre-amplifier and 146~mm for the main-amplifier) \cite{koechner06a}. For pulse energies larger than the stored energy of the amplifier, the gain is proportional to the population inversion $G\propto \Delta$ \cite{park18a,jeong17a}. In our case, for generating rectangular pulses we operate in an intermediate regime, which cannot be described by a simple expression. Therefore the intensity of the seed pulse must be carefully ramped up as the gain drops in both amplifiers. As expected due to the exponential relationship between the population inversion and gain, this effect is particularly apparent in the pre-amplifier; this can be seen in the first two rows of the 200~$\mu$s and 60~$\mu$s panels in Fig.~\ref{fig_time_profile}. With proper seed pulse shaping that takes the time-dependence of the temporal gain profile into account, any overshoot or spikes can be avoided to guarantee a temporally flat-top pulse. A smooth temporal profile is required to maintain narrow linewidths. The temporally flat-top pulse is one example of this, but our pulse-shaping method could clearly be modified to give other smooth temporal pulse shapes (e.g. Gaussian).

The current setup was used to generate arbitrary pulse lengths between 4 and 200~$\mu$s. We were able to obtain higher peak power for shorter pulses.  For example, with a 60~$\mu$s pulse we obtained up to 4~kW in the fundamental and 2~kW in the second harmonic. (The data depicted in Fig.~\ref{fig_time_profile} was obtained for a lower amplifier drive current to avoid optical damage of the LBO crystal under the described focusing condition.)

\begin{figure}[h!]
\centering\includegraphics[width=0.7\columnwidth]{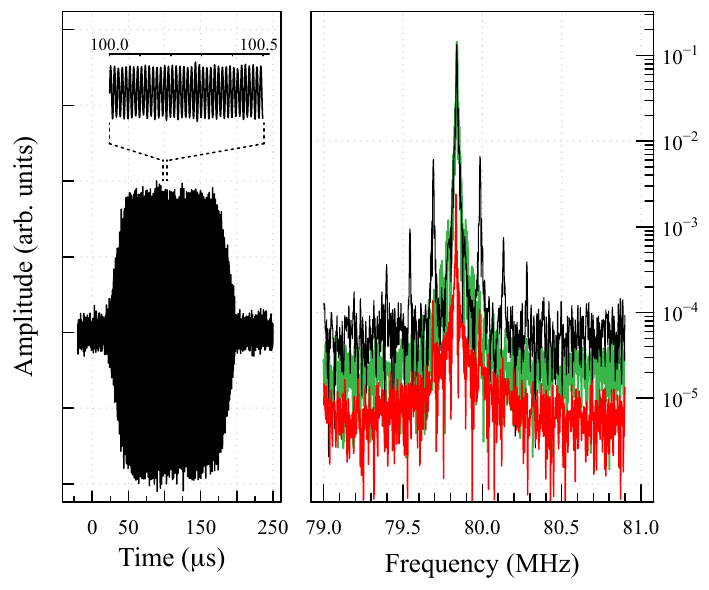}
\caption{{\it Left:} Beat-note signal arising from the superposition of the cw output of the master laser with the pulsed infrared light. The oscillation frequency of $\sim$80~MHz corresponds to the frequency of the accousto-optic modulator used to generate a seed pulse from the cw output. {\it Right:} Beat-note spectrum of the continuous wave seed (red) and amplified (black) laser pulse. The beat-note spectrum of a synthetic pulse is shown in green. }
\label{fig_beat}
\end{figure}

To determine the effect of the amplification on the linewidth of the output laser pulse, a beat-note measurement is performed by overlapping a fraction of the master oscillator beam (taken from the undiffracted beam after the AOM) with the amplified 1064 nm pulse on a fast photodiode. The resulting beat-note signal (Fig.~\ref{fig_beat}) shows the expected oscillation at the AOM drive frequency, $\approx 80$~MHz. Fig.~\ref{fig_beat} also displays the beat-note spectrum of the unamplified cw seed beam (red) and the amplified pulse (black). Because phase noise on the cw seed beam cannot affect this beat-note signal, we attribute the 4~kHz FWHM width of the cw beat-note to amplitude noise (either in the seed beam or our detection electronics). We attribute the sidebands at $\pm 150$~kHz from the carrier to amplitude modulation of the seed laser. The beat-note of the amplified beam shows a FWHM of $8$~kHz, indicating that the amplification does not broaden the linewidth of the master oscillator (specified to be less than 5 kHz) by more than a few kHz. 
Note that the sidebands on the cw spectrum are essentially reproduced in the amplified pulse spectrum. Sidebands resulting from the particular temporal envelope of the amplified pulse are estimated to appear at $\pm70$~kHz and with an amplitude $\sim$150 times smaller than the carrier frequency, as indicated by the green trace in Fig.~\ref{fig_beat}. The green trace was obtained as the Fourier transform of a simulated beat-note signal of equal S/N of a pulse with similar envelope and single frequency.

The second harmonic generation conversion efficiency with 2.5 kW power in the fundamental is predicted to be 52\% by an mlSNLO \cite{smith04a} calculation for a 50~mm long LBO crystal with a beam waist of 70~$\mu$m at the center. The observed energy-conversion efficiencies for the cw-seeded, 200~$\mu$s, and 60~$\mu$s pulses are 46\%, 50\%, and 53\%, respectively--in good agreement with the prediction.

\begin{figure}[h!]
\centering\includegraphics[width=\columnwidth]{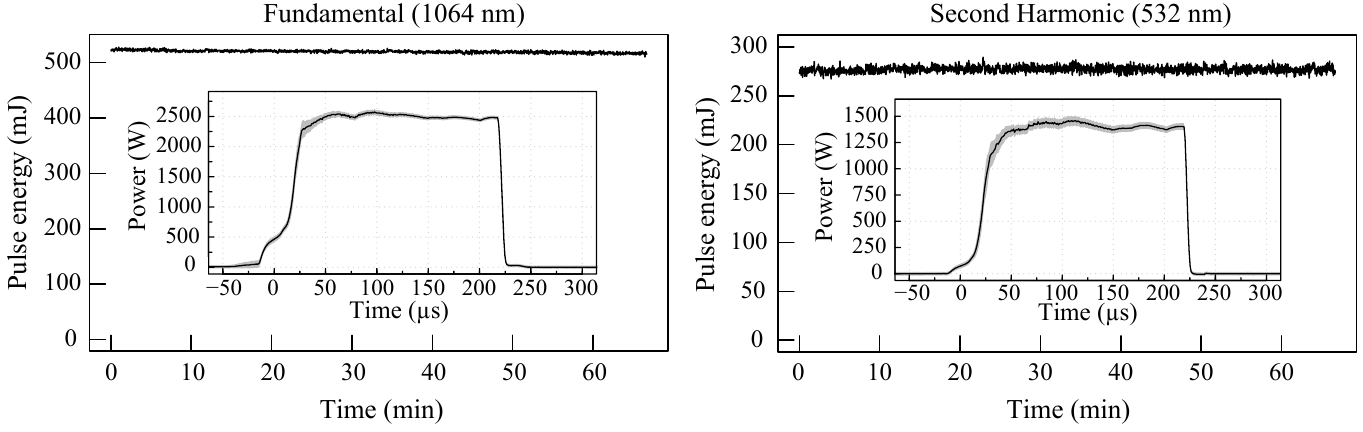}
\caption{Pulse energy of the fundamental frequency (1064~nm) and second harmonic (532~nm) versus time, over a 60~min interval. The insets show the mean temporal profile of the laser pulses and its 1$\sigma$ range of variation (in grey) over 60~min. }
\label{fig_stability}
\end{figure}

Fig.~\ref{fig_stability} shows the pulse energy for the fundamental frequency and the second harmonic over a 60 minute time interval, indicating a long-term stability of better than 0.5\% and 1.1\%, respectively. Also shown are the averaged temporal profiles of the laser pulses and the 1$\sigma$ ranges of deviation from the average over the same 60 minute interval. The individual pulse shapes are preserved to better than 1.8\% for the fundamental and 2.9\% for the SH over 60 minutes.

\begin{figure}[h!]
\centering\includegraphics[width=\columnwidth]{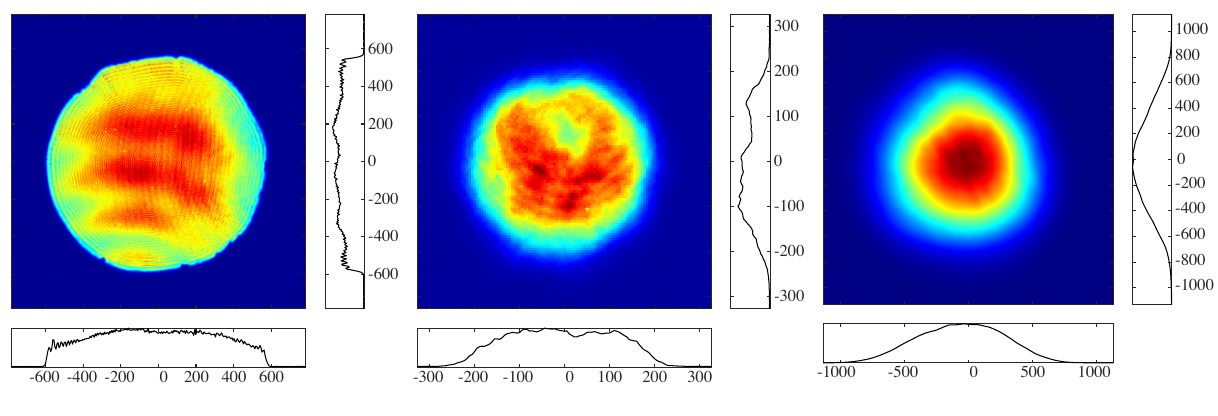}
\caption{Beam profiles of the 1064~nm seed beam (left), amplified 1064~nm (middle) and SH (right). Units are in $\mu$m and all profiles are taken at relay planes. }
\label{fig_beam_profile}
\end{figure}

Spatial profiles of the apertured seed, amplified fundamental, and SH beams are shown in Fig.~\ref{fig_beam_profile}. The amplification process leads to a degradation of the flat-top beam profile obtained using the aperture, as can be seen by comparing the first two panels in Fig.~\ref{fig_beam_profile}. To reach high conversion efficiency in the SH generation, nonlinear crystals with a length of several centimeters and a near confocally-focused input beam must be employed. The focused flat-top beam profile cannot be maintained under these conditions. With the beam focused at the center of the crystal, we obtain a SH beam with a Gaussian profile, as is apparent in Fig.~\ref{fig_beam_profile}. The M$^2$ for the SH beam was determined to be 1.41(14).

\section{Conclusion and Outlook}

We have demonstrated efficient amplification of the output of a narrow-bandwidth seed laser using a two-stage diode-pumped Nd:YAG amplification setup, reaching 2.6~kW at 1064~nm and 1.3~kW at 532~nm for a pulse length of 200~$\mu$s. A flat-top seed beam profile and long diode-pumping pulses resulted in a gain of 9600 and 16 in the double-passed pre-amplifier and main-amplifier, respectively. In combination with an acousto-optic modulator for pulse shaping of the cw master oscillator output, this enabled the generation of rectangular time-profile pulses with widths on the timescale of the fluorescence lifetime of the amplification medium. The single-frequency output with $<10$~kHz linewidth, excellent beam quality in the second harmonic, and stable, high peak powers realized here compares favorably to other, previously demonstrated systems with shorter pulse lengths, larger linewidth, higher complexity and larger number of gain modules \cite{meijer17a,yarrow07a,xu15a,bian16a,bian16b, guo16a, zong18a, zong19b}.  

This compact laser system provides desirable characteristics for potential use as a pump for tunable, high-power, narrow-linewidth long-pulse laser systems. Such lasers could in principle be based on optical-parametric oscillators \cite{devi17a, aoust14a, couvin16a}, or on amplification of weaker narrowband cw lasers using gain in media such as Ti:sapphire \cite{sprecher13a, xu17a, zong19a}, dye \cite{baving82a}, or nonlinear crystals (via optical parametric amplification). The smooth temporal profile and the complete absence of laser spiking in the pump source will minimize the danger of optical damage and provide steady conditions in the pumped medium so as to avoid transient thermal lenses \cite{lausten03a,planchon08a}. The kHz linewidth of the described pump laser is a prerequisite to achieve narrow linewidths in optical-parametric oscillators and amplifiers.

We anticipate that the laser system described here will stimulate further work in the direction of using stimulated light forces in molecular laser cooling experiments.

\section*{Acknowledgments}
The authors thank Jai-Min Choi for technical support. This work was supported by ARO DURIP grant W911NF-12-1-0475.

\section*{Disclosures}

\noindent The authors declare no conflicts of interest.

\section*{Data availability} Data underlying the results presented in this paper are not publicly available at this time but may be obtained from the authors upon reasonable request.

%%%%%%%%%%%%%%%%%%%%%%% References %%%%%%%%%%%%%%%%%%%%%%%%%

%%%%%%%%%% If using BibTeX:
\bibliography{lit}

\begin{thebibliography}{10}
\bibitem{lumentum}
{The master oscillator laser was a Lumentum 126-1064-100.}
\end{thebibliography}

\end{document}